\documentclass{iopjournal1}

\usepackage{verbatim}
\usepackage{graphicx}
\usepackage{multirow}
\usepackage{subfigure}
\usepackage{amsfonts}
\usepackage{amsmath}
\usepackage{amsbsy}
\usepackage{color}
\usepackage{bm}
\usepackage{esint}
\usepackage{cleveref}
\usepackage{pifont}
\usepackage{ulem}
\usepackage{orcidlink}

\newcommand{\doiurl}[2]{\href{https://doi.org/\detokenize{#1}}{#2}}
\renewcommand{\vec}[1]{\mbox{\boldmath$\mathrm{#1}$}}
\let\sb=_ \catcode`\_=\active \def_#1{\ensuremath \sb{\rm#1}}
\newcommand{\coloredsubref}[2]{\hyperref[#1]{\ref*{#1}#2}}

\begin{document}

\articletype{Paper} 
 
\title{Spin-Based True Random Number Generation Enabled by Voltage-Amplified Quantum Fluctuations}

\author{Jie Zheng$^1$\orcidlink{0009-0009-4528-5525}, Jiyong Kang$^{1,4}$\orcidlink{0009-0007-4023-7212}, Zheng Zhu$^5$, Di Wu$^5$, Yuesheng Li$^5$, Dongxing Yu$^{1,*}$\orcidlink{0000-0002-3622-5063}, Jiayong Wang$^{5,*}$\orcidlink{0009-0003-6687-2530}, Hongxing Xu$^{6,7,*}$\orcidlink{0000-0002-1718-8834} and Chenglong Jia$^{1,2,3,*}$\orcidlink{0000-0001-7673-050X}}

\affil{$^1$ School of Physical Science and Technology, Lanzhou University, Lanzhou 730000, China.}

\affil{$^2$ Lanzhou Center for Theoretical Physics, Key Laboratory of Quantum Theory and Application of MoE, Lanzhou University, Lanzhou 730000, China.}

\affil{$^3$ Key Laboratory of Theoretical Physics of Gansu Province, Gansu Provincial Research Center for Basic Disciplines of Quantum Physics, Lanzhou University, Lanzhou 730000, China.}

\affil{$^4$ Songshan Lake Materials Laboratory Dongguan, Guangdong 523808, China}

\affil{$^5$ Institute of Integrated Circuits, Shanghai University, Shanghai, 201800, China.}

\affil{$^6$ Henan Academy of Sciences, Zhengzhou, 450046, China.}

\affil{$^7$ Wuhan University, Wuhan, 430072, China.}

\affil{$^*$ Author to whom any correspondence should be addressed.}

\email{yudx@lzu.edu.cn, wangjiayong@shu.edu.cn, hxxu@hnas.ac.cn and cljia@lzu.edu.cn}


\begin{abstract}
We investigate spin quantum-fluctuation effects that originate from the Heisenberg uncertainty principle during the dynamical cycle of disentanglement, entanglement, and re-disentanglement between itinerant electrons and localized magnetic moments mediated by the {\it s–d} exchange interaction. Beyond conventional deterministic spin-transfer torque, we analyze an intrinsic mechanism that transfers spin quantum fluctuations to a nanomagnet. By extending the Landau–Lifshitz–Gilbert equation to incorporate both quantum and thermal stochastic fields, we identify a temperature regime in which quantum fluctuations dominate the magnetization dynamics. We further show that voltage-controlled magnetic anisotropy exponentially amplifies spin quantum fluctuations, enabling binary readout through magnetoresistance in magnetic tunnel junctions. These findings provide a microscopic framework for fluctuation-driven spin dynamics and outline a device-level pathway toward spin-based quantum true random number generation.
\end{abstract}

\section{\label{sec:level1} Introduction}
       Random numbers are essential for cryptography, data encryption, privacy protection, and stochastic simulations \cite{ma2016a,bykovsky2018,saini2022,miszczak2013}. Depending on their origin, random number generators (RNGs) are classified as pseudo-random number generators \cite{james1990}, which rely on deterministic algorithms, or true random number generators (TRNGs) \cite{fu2021,kang2025}, whose randomness arises from intrinsically unpredictable physical processes. Owing to their physical basis, TRNGs offer superior security and reliability.

       Quantum true random number generators (QTRNGs) \cite{herrero-collantes2017} harness the intrinsic indeterminacy of quantum mechanics, with entropy sources including quantum phase noise \cite{qi2010}, vacuum fluctuations \cite{xu2012}, random path selection \cite{jennewein2000}, and quantum tunneling \cite{zhou2019}. These mechanisms yield high-quality randomness for applications demanding exceptional security, such as quantum communication and financial encryption.

       Among all sources of quantum randomness, the intrinsic uncertainty prescribed by the Heisenberg principle constitutes a universal and fundamental physical mechanism \cite{dirac1981}. In magnetic systems, spin quantum fluctuations originate from the noncommutativity of angular momentum operators,
       $[\hat{J}_i,\hat{J}_j]=i\epsilon_{ijk}\hat{J}_k$ with $\hbar=1$, leading to the uncertainty relation $\Delta J_i\Delta J_j\geq|\langle\hat{J}_k\rangle|/2.$ These quantum fluctuations are especially pronounced in spin-1/2 systems and nanoscale magnetic structures. Experiments by Zholud {\it{et al.}} demonstrated that quantum and thermal spin fluctuations contribute differently to magnetization relaxation dynamics, with quantum fluctuations dominating at low temperatures \cite{zholud2017a,zhang2017a}.
 
       In this work, we investigate the dynamics of quantum fluctuations within the {\it s–d} exchange interaction between itinerant $s$-electrons and localized $d$-electrons in ferromagnetic metals. By tracing the complete disentanglement–entanglement–redisentanglement cycle between these degrees of freedom, we reveal how spin quantum fluctuations evolve during angular momentum transfer beyond the deterministic spin-transfer torque (STT) mechanism \cite{berger1996,slonczewski1996}. Extending the Landau–Lifshitz–Gilbert (LLG) equation to incorporate stochastic quantum and thermal fields, we identify the temperature regime in which quantum fluctuations dominate the magnetization dynamics. Furthermore, the voltage-controlled magnetic anisotropy (VCMA) effect \cite{kang2017a,dai2022,matsumoto2023b} enables exponential amplification of these fluctuation-induced dynamics, which are subsequently converted into binary stable states in magnetic tunnel junctions (MTJs) \cite{chen2022}. The tunneling magnetoresistance (TMR) effect \cite{WOS:000181101000005} provides direct electrical readout of these quantum-originated signals. Our findings establish a microscopic framework for fluctuation-driven spin dynamics and suggest a fundamental route toward spin-based quantum true random number generation.

\section{\label{sec:level2} Transfer of spin quantum fluctuations}

       \begin{figure}
       \centering
        \includegraphics[width=0.75\textwidth]{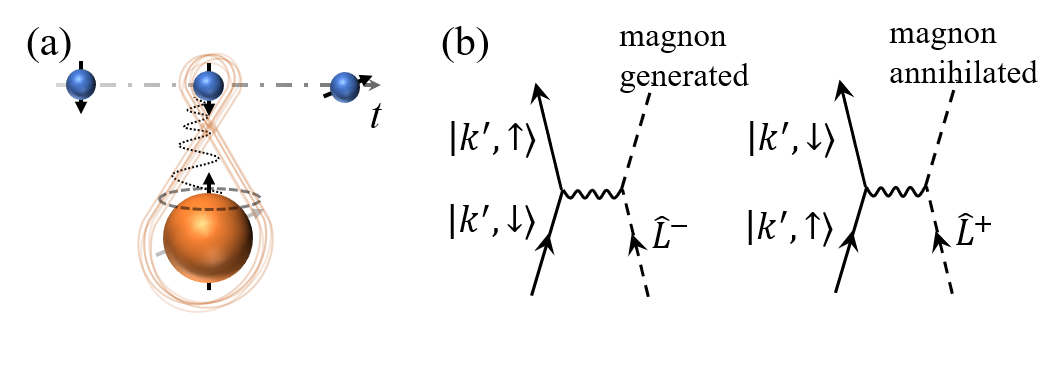}
       \caption{Quantum-state evolution during {\it s–d} scattering. (a) The itinerant $s$-electron and localized moment are initially and finally uncorrelated, but become entangled within the interaction region: a spin-down electron and a spin-up local moment form a high-energy spin-singlet state under strong ferromagnetic {\it s–d} exchange. (b) Feynman diagrams of {\it s–d} scattering, where electron spin flips correspond to magnon creation or annihilation, transferring both angular momentum and spin quantum fluctuations. }
       \label{fig1}
       \end{figure}

       We consider the dynamic evolution of an incident conduction $s$-electron spin ($\hat{\vec{s}}$) interacting with a localized magnetic moment ($\hat{\vec{L}}$) through the short-range {\it s-d} exchange coupling \cite{yosida1996theory}, 
	\begin{equation}
		H_{sd} =- \lambda\hat{\vec{s}}\cdot\hat{\vec{L}}  \label{eq:1},
	\end{equation}
	where $\lambda$ denotes the ferromagnetic exchange strength. Before scattering, the two spins occupy uncorrelated states, so that the total incoming and outgoing states can be written as
	 \begin{equation}
	 \lvert\psi\rangle_{in} =\lvert \psi\rangle^{s}_{i}\otimes \lvert \psi\rangle^{d}_{i}, \qquad 
	 \lvert \psi\rangle_{out} =\mathcal{S} (\lvert \psi\rangle^{\rm s}_{i}\otimes \lvert \psi\rangle^{\rm d}_{i}),
	 \label{eq:2}
	 \end{equation}
	with $\mathcal{S}$ being the unitary scattering matrix. Inside the interaction region, the spin dynamics is naturally expressed in the eigenbasis of total angular momentum $\hat{\vec{J}} =  \hat{\vec{L}}+\hat{\vec{s}}$, 
	 \begin{equation}
	 \lvert J, J_z \rangle = C_{\uparrow} \lvert m, \tfrac{1}{2} \rangle + C_{\downarrow} \lvert m+1, -\tfrac{1}{2} \rangle ,
	 \label{eq:3}
	 \end{equation}
	where $C_{\uparrow, \downarrow} $ are the Clebsch-Gordan coefficients \cite{sakurai2020modern}. It demonstrates that all but the fully polarized $\lvert J, \pm J \rangle$ channels form entangled superpositions.
	Thus, the scattering event finishes a cycle of disentanglement $\rightarrow$ entanglement $\rightarrow$ redisentanglement, enabling transfer of both angular momentum and intrinsic spin quantum fluctuations figure~\ref{fig1}.

	 The minimal quantum fluctuations of $s$- and $d$-electron spins satisfy the Heisenberg relations: $\Delta s_x\Delta s_y=\lvert\langle\hat{s}_z\rangle\lvert /2$ and $\Delta L_x\Delta L_y=\lvert\langle\hat{L}_z\rangle\lvert/2$, 
	while $\hat{\bf{J}}^2$ and $\hat{J}_z$ remain conserved throughout the scattering process.
	During the final redisentanglement stage, the fluctuation transfer obeys
	\begin{equation}
		\Delta s_z+\Delta L_z=0,
		\label{eq:4}
	\end{equation}
	allowing quantification via either $\Delta s_z=\langle\hat{s}_z\rangle_{out}-\langle\hat{s}_z\rangle_{in}$ or $\Delta L_z=\langle\hat{L}_z\rangle_{out}-\langle\hat{L}_z\rangle_{in}$. 

       For a general incoming state,
	 \begin{eqnarray}
	 \lvert\psi\rangle_{in}
	  = \left(a\lvert k,\tfrac{1}{2}\rangle+b\lvert k,-\tfrac{1}{2}\rangle\right)\otimes\sum_{m=-l}^{l}f(m)\lvert l,m\rangle,\label{eq:5}
	 \end{eqnarray}
	 with  $k$ being the wave vector of the incident $s$-electron,  $\lvert a\lvert^2+\lvert b\lvert^2=1$ and $\sum_{m} f^{*}(m)f(m)=1$, 
	 we treat the conduction electrons as an effective environment, 
	 and introduce the Kraus operator \cite{kraus1983a}
	 \begin{equation}
	 \mathcal{K}_{k,s;k',s'}\equiv\langle k,s\lvert\mathcal{S}\lvert k',s'\rangle. \label{eq:6}
	 \end{equation}	 
	 The outgoing state then becomes
	 \begin{equation}
	    \lvert\psi\rangle_{out} 
		=\sum_{s,k'}\sum_{m'}\left(\lvert k',s\rangle\otimes\lvert l,m'\rangle\right)\langle l,m'\lvert\left(a\mathcal{K}_{k',s;k,+}+b\mathcal{K}_{k',s;k,-}\right)\lvert\psi\rangle_{i}^{\rm d}.\label{eq:7}
	 \end{equation}
        The corresponding change in $\langle \hat{L}_z \rangle$ is
        \begin{equation}
	     \Delta L_z 
         =\sum_{m}\left\{(\xi_{l,m}^{+}\lvert a\rvert^2 -\xi_{l,m}^{-}\lvert b\rvert^2) \sin^2 \phi \lvert f(m)\rvert^2+\Re{[\zeta_{l,m}a^{*}bf^{*}(m)f(m+1)]}\right\},\label{eq:8}
	 \end{equation}
	 where $\xi_{l,m}^{\pm}$ and $\zeta_{l,m}$ depend on the Clebsch-Gordan coefficients, and $\phi$ is the relative phase between scattering channels. The first terms reflect individual $J$-channel contributions, while the last term arises from channel interference.

        If the localized magnetic moment initially occupies $\lvert \psi_i^{\rm d} \rangle = \lvert l,m \rangle$ with magnon number $N=l-m \ll l$ \cite{zholud2017a,auerbach1994interacting,urazhdin2004a}, then
	 \begin{equation}
		\Delta N = - \Delta L_z \approx \left[-\frac{\lvert a\rvert^2}{l}N+\frac{\lvert b\rvert^2}{l}N+\frac{\lvert b\rvert^2}{l}\right]\sin^2 \phi.\label{eq:9}
	 \end{equation}
	 Here, terms proportional to $N$ correspond to stimulated magnon absorption/emission, whereas the $N$-independent term represents spontaneous magnon emission, directly signaling spin quantum fluctuations associated with spin flips of the $s$-electron between two energy-levels  with $J=l \pm 1/2$.

        Several specific cases follow immediately. For $m=l$ and $(a,b)=(1,0)$, $\vec{s}$ is parallel to $\vec{L}$, yielding $\Delta N=0$. For $(a,b)=(0,1)$, they are antiparallel, giving $\Delta N\propto 1/l$, entirely from spontaneous magnon emission. For a $s_z$-unpolarized incident electron ($a = b = 1/\sqrt{2}$), $\Delta N\propto 1/(2l)$, where stimulated processes cancel while spontaneous magnon emission remains. Therefore, the transfer of spin quantum fluctuations is universal, but its magnitude depends on the relative orientation between the incident electron spin $\vec{s}$ and the local magnetic moment $\vec{L}$, reaching its maximum (minimum) when $\vec{s}$ and $\vec{L}$ are antiparallel (parallel). 
	 
	 Extending this picture to continuous scattering from a spin-polarized current with electron interval $\tau$ \cite{wang2012,wang2013a}, and noting that {\it s–d} scattering occurs on femtosecond timescales \cite{jia2014}, which is much faster than the relaxation/precession of magnetic moment $\vec{m}$,
       the density-matrix dynamics of the local moment, $\rho^L$, evolves as
	 \begin{equation}
	  \frac{\partial }{\partial t}\rho^L=\frac{\rho^L_{out}-\rho^L_{in}}{\tau} .\label{eq:10}
	 \end{equation}
	Using the $\mathcal{P}$-representation of the spin-coherent-state \cite{zhang1990,narducci1975}, one has
		$\rho^L(t)=\int d\Omega \mathcal{P}^L(\Omega,t)\lvert L,\Omega\rangle\langle L,\Omega\rvert$.
	Tracing out the electron degrees of freedom yields a Fokker–Planck equation,
	\begin{equation}
	\frac{\partial\mathcal{P}^L}{\partial t}
	=-\boldsymbol{\nabla} \cdot [\vec{T}\mathcal{P}^L]+\nabla^2[\mathcal{D}\mathcal{P}^L].
	\label{eq:11}
	\end{equation}
	with drift
       $\vec{T}=\mathcal{A}(\vec{m}\times\vec{s})\times\vec{m}+\mathcal{B}(\vec{m}\times\vec{s})$,
       and diffusion
       $\mathcal{D}=\frac{\mathcal{A}}{2l+1}(1-\vec{s}\cdot \vec{m})$.
       The drift term reproduces damping-like and field-like STT via stimulated processes, whereas the diffusion term originates from spontaneous magnon emission and encodes spin quantum fluctuations. The coefficients $\mathcal{A}$ and $\mathcal{B}$ depend on {\it s–d} exchange and scale with current.

       It should be noted that although $\mathcal{P}^L(\Omega,t)$ is normalized, it may become negative in strongly nonclassical regions \cite{zhang1990,narducci1975}. Direct interpretation as a classical probability $P^L(\theta,\phi,t)$ is therefore problematic. To incorporate fluctuation strength into the LLG formalism while preserving classical interpretability, we adopt a WKB approximation in the large-$l$ limit.

       \begin{figure}
       \centering
        \includegraphics[width=0.35\textwidth]{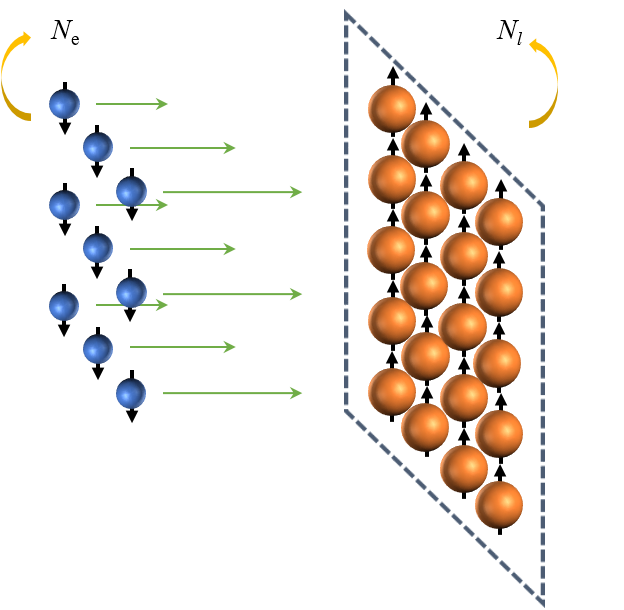}
       \caption{Scattering between an incident electron current and a magnetic film. Within the time interval $\tau$, $N_e$ spin-polarized electrons are injected into the film, where they undergo scattering with $N_l$  local magnetic moments.}
       \label{fig2}
       \end{figure}

\section{\label{sec:level3} Stochastic fields}    

       For a two-dimensional electron current incident on a single-domain magnetic film, the {\it s–d} exchange strength is effectively renormalized by ferromagnetic interactions among local moments, modifying mainly the phase $\phi$ in Eq.~\eqref{eq:8}. Individual scattering events remain governed by the Hamiltonian in Eq.~\eqref{eq:1}. Averaging over the $N_e$ electrons and $N_{\it l}$ moments within the independent-scattering approximation, as shown in figure~\ref{fig2}, yields a stochastic LLG description, in which the net {\it s–d} interaction appears as a deterministic STT and a quantum stochastic field, representing the coherent and fluctuating channels of angular-momentum transfer.

       The magnetization dynamics is described by the stochastic LLG equation \cite{brown1963,garcia-palacios1998},
	 \begin{equation}
	 \frac{d\vec{M}}{dt}=-\gamma\vec{M}\times(\vec{H}_{\text{eff}}+\vec{h})+\boldsymbol{\Gamma}_s-\frac{\gamma\alpha}{M_s}\vec{M}\times[\vec{M}  \times(\vec{H}_{\text{eff}}+\vec{h})],\label{eq:12}
	 \end{equation}
	 where $\vec{M}$ is the magnetization with magnitude $M_s= \gamma_0 N_{\it l} l \hbar /V$, $\alpha$ is the Gilbert damping, $\gamma=\frac{\gamma_0}{1+\alpha^2}$ is the gyromagnetic ratio, $\vec{H}_{\text{eff}}$ is the effective field \cite{yu2021}, $\vec{h}$ is the random field, and $\vec{\Gamma}_s$ denotes the STT.

       The random field is taken to be Gaussian \cite{brown1963}
	 \begin{equation}
	 \langle h_i\rangle=0,\;\;\langle h_i(t)h_j(t')\rangle=2D'\delta_{ij}\delta(t-t'),\label{eq:13}
 	 \end{equation}
	 with $h_i(t)=h_{T,i}(t)+h_{Q,i}(t)$ and the fluctuation strength $D'= D_T+D_Q$. Here, thermal ($\vec{h}_T$) and quantum ($\vec{h}_Q$) fluctuations are treated as statistically independent.  The corresponding quansi-classical Fokker–Planck equation for $P^L(\theta,\phi,t)$ reads then \cite{risken1989fokker}:
	 \begin{equation}
	 \frac{\partial P^L(\theta,\phi,t)}{\partial t}=-\boldsymbol{\nabla}\cdot\left[\left(\gamma\alpha\vec{H}_{\text{eff}}+\frac{\boldsymbol{\Gamma}_s}{M_s}\right)P^L(\theta,\phi,t)\right]+\nabla^2\left[\gamma^2(1+\alpha^2)D'P^L(\theta,\phi,t)\right].\label{eq:14}
 	 \end{equation}

        With an average scattering probability $\epsilon=N_e/N_{\it l}$ over time $\tau$, the STT derived from the quantum master equation takes the form
	 \begin{equation}
	 \vec{\Gamma}_s=\epsilon M_s\vec{T}=a_M\vec{M}\times(\vec{M}\times \vec{p})+b_M\vec{M}\times\vec{p}, \label{eq:15}
	 \end{equation}
	 where $\vec{p}$ is the incident-electron spin-polarization and
	 \begin{equation} 
	   a_M=-\frac{N_e\sin^2\phi}{N_{\it l}(2l+1)M_s\tau},~~ b_M=-\frac{N_e\sin\phi\cos\phi}{N_{\it l}(2l+1)\tau}.\label{eq:16}
	 \end{equation}
        Given that the electron current density $I_e={eN_e\chi}/{\tau}$ with $\chi$ being the transmission probability, the drift term in Eq. (\ref{eq:11}) reproduces the Slonczewski torque \cite{berger1996,slonczewski1996}. 

	 In the absence of STT ($\vec{\Gamma}_s=0$) and for purely thermal noise ($D'=D_T$), the stationary distribution reduces to the Boltzmann form, which yields the standard thermal fluctuation strength \cite{li2004}
	 \begin{equation}
	 D_T=\frac{\alpha}{1+\alpha^2}\frac{k_B T}{\gamma M_sV}=\frac{\alpha k_B T}{\gamma_0 M_sV},\label{eq:17}
	 \end{equation}
	 where $k_B$ is the Boltzmann constant, and $T$ is the temperature. 
	
	 Identifying the quasi-probability $\mathcal{P}^L$ with $P^L$ and averaging over $N_{\it l}$ local moments gives the nanomagnet’s quantum fluctuation strength
	 \begin{equation}
	 D_Q \approx \frac{(1+\alpha^2)N_e\sin^2\phi}{\gamma_0^2N_{\it l}^2(2l+1)^2\tau}(1-\frac{\vec{M}}{M_s}\cdot\vec{p}).\label{eq:18}
	 \end{equation}
        Thus, $D_Q$ increases with current density $I_e$, scales as $1/[N_{\it l}(2l+1)]^2$, and is minimized (maximized) when $\vec{p}$ is aligned (anti-aligned) with $\vec{M}$.

        Equating $D_T$ and $D_Q$ yields the quantum–classical crossover temperature
	 \begin{equation}
	 T_Q = \frac{\gamma_0 M_sV D_Q}{\alpha k_B}. \label{eq:19}
	 \end{equation}
	 When $T < T_Q$, quantum fluctuations prevail; otherwise, thermal fluctuations dominate \cite{wang2013a}. Clearly, systems with lower damping $\alpha$ exhibit a higher crossover temperature.

\begin{figure}[t]
       \centering
        \includegraphics[width=0.85\textwidth]{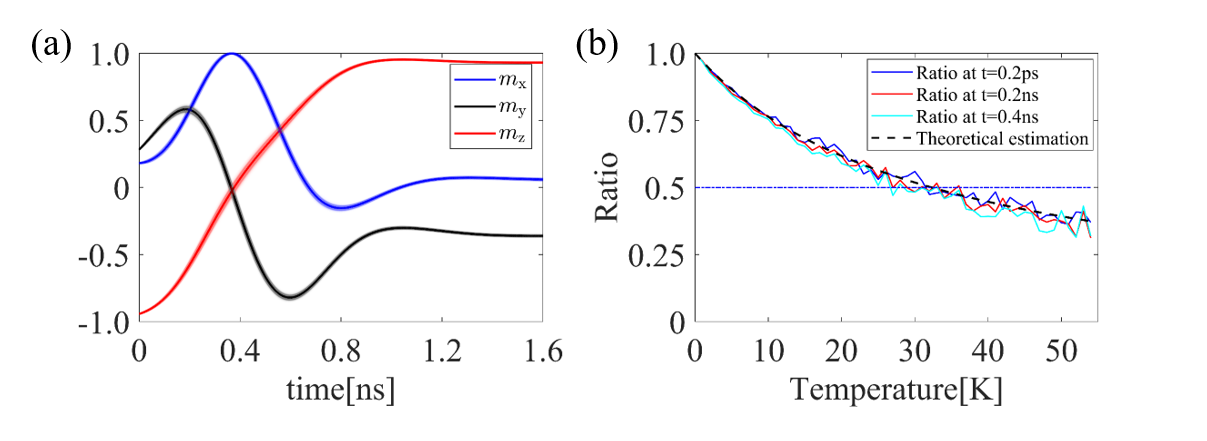}
        \caption{Stochastic magnetization dynamics and     quantum-fluctuation ratio.
        (a) Time evolution of the average magnetization components (solid lines) together with individual stochastic trajectories (transparent lines) at $T = 10\,\mathrm{K}$.
        (b) Temperature dependence of the quantum-fluctuation ratio sampled at $t = 0.2\,\mathrm{ps}$ (blue solid line), $0.2\,\mathrm{ns}$ (red solid line), and $0.4\,\mathrm{ns}$ (cyan solid line) in the simulations, compared with the theoretical prediction (black dashed line).}
       \label{fig3}
       \end{figure}

        To quantify the relative contribution of quantum fluctuations to the stochastic dynamics, we perform macrospin simulations. The circular nanomagnet lies in the $xy$-plane with saturation magnetization $M_s=1200$ kA/m, radius $r=10$ nm, thickness $t_f=1.1$ nm, and the damping $\alpha=0.001$. The initial magnetization direction is $\Omega_0=(2.8,1.0)$, and an external magnetic field of magnitude $B=0.05$ T is applied along $\Omega_0$. A spin-polarized current  of amplitude 60 $\mu$A, with the polarization $\mathbf{p}=(0,0,1)$, is injected along the $z$-direction. For each temperature, we perform 1000 stochastic realizations. The representative trajectories at T=10 K are shown in figure~\coloredsubref{fig3}{(a)}.

        The fluctuation intensity is characterized by the standard deviation of the stochastic trajectories
        \begin{equation}
        \delta N=1-\sqrt{1-(\delta\theta^2+\sin^2\bar{\theta}\delta\phi^2)}\label{20}
        \end{equation}
        where $\bar{\theta}$ is the mean polar angle, and $\delta\theta_{tot}$ and $\delta\phi$ are the corresponding standard deviations. We define the quantum-fluctuation ratio as
        \begin{equation}
        \mathrm{Ratio} = \frac{\delta N_{tot}-\delta N_T}{\delta N_{tot}}, \label{21}
        \end{equation}
        where $\delta N_{tot}$ and $\delta N_{T}$ denote the total and purely thermal fluctuation amplitudes, respectively.
        The theoretical estimate, $\frac{D_Q(\Omega_0)}{D_Q(\Omega_0)+D_T}$, is in good agreement with the numerical results shown in figure~\coloredsubref{fig3}{(b)}. From this comparison, we identify a quantum–classical crossover temperature $T_Q =32$ K. 

        For an MTJ, the free-layer switching probability in the presence of both fluctuation sources is  \cite{brown1963,li2004}
	 \begin{equation}
	 P_{sw}(t)=1-\exp\left[-\frac{t}{\tau_0}\exp\left(-\frac{E_b}{k_BT^{\star}}\right)\right],\label{eq:22}
	 \end{equation}
	 where $\tau_0$ is the attempt time, $E_b$ is the energy barrier between the low-resistance parallel (P) state and the high-resistance antiparallel (AP) state. The effective temperature $T+T_Q$ incorporates both thermal and quantum fluctuations.

        Our estimate of quantum fluctuations is obtained by averaging the microscopic fluctuations of individual localized moments. This approach differs fundamentally from the macrospin model, where only coherent quantum fluctuations of the collective magnetization are retained while incoherent magnons are neglected \cite{urazhdin2004a,wang2013a}. In contrast, our model begins with nearly free magnetic moments, includes the quantum fluctuations generated during scattering, and averages their collective dynamics assuming uniform precession in the single-domain limit. Consequently, our approach captures only fully incoherent fluctuations, whereas partially coherent or fully coherent (macrospin) contributions are not included. Both approaches therefore underestimate the full quantum-fluctuation strength in realistic nanomagnets.

       \begin{figure}
       \centering
        \includegraphics[width=0.85\textwidth]{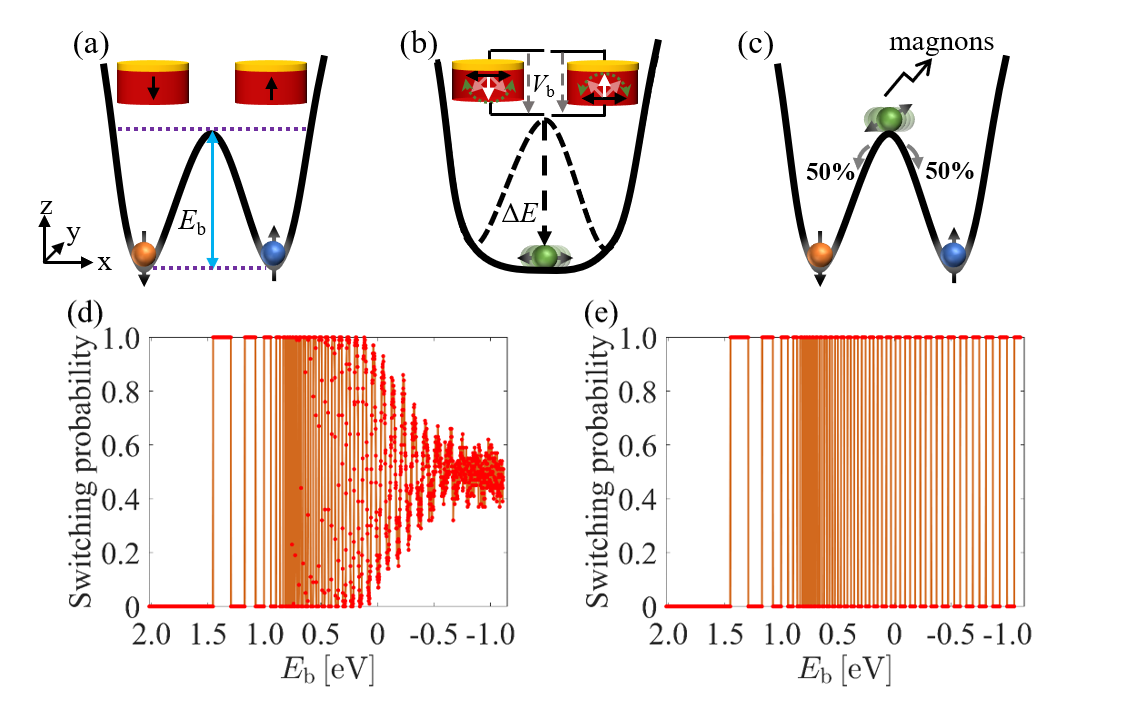}
        \caption{VCMA-based amplification of spin quantum fluctuations.
	 (a) A perpendicularly magnetized free layer has two stable states separated by an energy barrier $E_b$.
	 (b) A VCMA voltage pulse lowers $E_b$, tilting the magnetization toward the $\pm x$ direction.
	 (c) After voltage removal, an $s_x$-polarized current injects spin quantum fluctuations, leading to stochastic relaxation into either state with nearly equal probability.
	 For $E_b=0$, the magnetization aligns with the applied in-plane field along $\pm x$.
	 Simulations start from $\vec{M}/M_s=(\sin(\pi/180),0,\cos(\pi/180))$, with a VCMA pulse of duration $T_1=20$ ns followed by an $s_x$-polarized current pulse ($I_e=2$ mA, $T_2=115$ ps, $\chi=1$).
	 Panels show $\vec{H}_x=(30,0,0)$ kA/m (d) and $(-30,0,0)$ kA/m (e). Each data point averages over 100 stochastic realizations.}
       \label{fig4}
       \end{figure}

\section{\label{sec:level4} Amplification of spin quantum fluctuations by VCMA }
        Because quantum fluctuations are intrinsically weak, they alone generate resistance variations far below the sensitivity of MTJ measurements. To render these fluctuations observable, we employ the VCMA effect figure~\ref{fig4}, which converts small magnetization perturbations into stochastic switching events, analogous to avalanche amplification in quantum-tunneling diodes \cite{zhou2019}. The voltage-dependent energy barrier is \cite{kang2017a}
	 \begin{equation}
	 E_b(V_b)=\left[K_i -{\xi_v V_b\over t_{ox}}-{\mu_0M_s^2\over 2}(N_z-N_{x,y})t_f\right]A,\label{eq:23}
	 \end{equation}
	 with $K_i$ the interfacial perpendicular anisotropy, $\xi_v$ is the VCMA coefficient, $t_{ox}$ is the oxide thickness, $N_z$ and $N_{x,y}$ are the demagnetization factors, $t_f$ the free-layer thickness, and $A$ the device cross-sectional area.

	 To illustrate the amplification mechanism, we perform macrospin simulations. The circular free layer lies in the $xy$-plane, and electrons are injected along $z$. We choose $M_s=1200$ kA/m, radius $r=40$ nm, thickness $t_f=1.1$ nm, localized spin $l=2$, and demagnetization factors $N_z=0.96$ and $N_{x,y}=0.02$. The interfacial anisotropy is set to $K_i =1$ mJ/m$^2$, the VCMA coefficient to $\xi_v=70$ fJ/(V·m), the oxide thickness to $t_{ox}=1.4$ nm, and the damping to $\alpha=0.05$. The phase factors describing the different entanglement modes in scattering are taken as $\sin^2 \phi=0.81$ and $\sin \phi \cos\phi =0.39$. 
	
	 The magnetization in the free layer is governed by the energy barrier between its states figure~\coloredsubref{fig4}{(a)}\cite{igarashi2024a}. A voltage pulse transiently lowers the energy barrier via VCMA while an in-plane field tilts the magnetization figure~\coloredsubref{fig4}{(b)}. Immediately after the pulse is removed, a spin-polarized current with $\vec{p}=(-1,0,0)$ injects quantum fluctuations into the free-layer dynamics figure~\coloredsubref{fig4}{(c)}. The MTJ then relaxes into the P or AP state, thereby converting quantum-induced fluctuations into discrete switching outcomes.
	  
        Figure~\ref{fig4} shows that the switching probability oscillates as the barrier is tuned, reflecting transient transitions between metastable configurations. When the free-layer magnetization is antiparallel to the current polarization, $D_Q$ is maximized and the switching probability approaches $50\%$ figure~\coloredsubref{fig4}{(d)}, whereas parallel alignment suppresses $D_Q$ and yields deterministic relaxation figure~\coloredsubref{fig4}{(e)}. This behavior directly confirms the angular dependence $D_Q\propto (1-\frac{\bf{M}}{M_s}\cdot\vec{p})$. For partially polarized currents, the minimum fluctuation amplitude remains finite \cite{wang2013b,wang2020c}.

        Treating quantum-induced MTJ relaxations as independent events, the number of switching outcomes in $N_t$ trials is binomially distributed \cite{zhou2019,caratozzolo2025},
	 \begin{equation}
	 P_{N_t}(k)=C_{N_t}^k p_{sw}^k(1-p_{sw})^{N_t-k}, 
	 \label{eq:24}\end{equation}
	 where $p_{sw}$ is the switching probability in a short interval $\Delta t$. By tuning the VCMA pulse amplitude and duration, $p_{sw}$ can be stabilized and reproducibly controlled, establishing a practical route to MTJ-based QTRNG operation.

\section{Conclusion} 

        Unlike thermal fluctuations, which arise from environmental agitation, spin quantum fluctuations, rooted in the Heisenberg uncertainty principle, persist at all temperatures and dominate magnetization dynamics in the low-temperature regime. Building on this property, the extended magnetodynamic framework developed in this work—incorporating both quantum and thermal random fields—provides a unified description of how multiple stochastic sources jointly shape magnetization dynamics. The temperature-dependent crossover between quantum- and thermal-dominated regimes further points to the possibility of temperature-tunable QTRNGs, in which the statistical characteristics of the output can be directly engineered through thermal control.

	 At the device level, we have proposed a practical approach for amplifying weak spin quantum fluctuations. The VCMA mechanism is used to convert small perturbations of the magnetization into stochastic state-switching events, while the TMR effect of MTJs enables high–signal-to-noise readout of these switching outcomes. Together, these effects overcome the longstanding challenge that quantum random signals are easily overwhelmed by classical noise during macroscopic measurements \cite{chen2022,ng2023,zhang2024}, thereby offering a feasible route toward experimentally accessing high-purity quantum randomness.

	Despite these advantages, several challenges remain before practical implementation. Realistic devices require systematic evaluation of additional noise sources, including quantum tunneling current fluctuations, shot noise, material defects, interface roughness, and other extrinsic perturbations, and their impact on the quality of the extracted random bits. Moreover, integrating the proposed mechanism with CMOS-compatible fabrication processes, while satisfying the high-throughput requirements of quantum cryptography and secure communication systems, represents an important direction for future research. Addressing these challenges will be essential for translating the present physics into scalable, robust, and commercially viable QTRNG hardware.

\ack{This work was supported by the National Natural Science Foundation of China (Grant Nos. 12574123, 12174164, 12247101, 12474114, and 12204497), and the 111 Project under Grant No. B20063.}

\data{The data that support the findings of this article are not publicly available. The data are available from the authors upon reasonable request.}

\normalem 

\end{document}